\documentclass{aa}
\usepackage{graphicx,natbib,color}

\bibpunct{(}{)}{;}{a}{}{,}

\newcommand{\ignore}[1]{}

\newcommand{\kpc}{\mathrm{kpc}}

\newcommand{\cl}[1]{#1}
\newcommand{\gc}[1]{#1}

\newcommand{\amd}[1]{#1}

\newcommand{\genco}[2]{#2}
\newcommand{\speco}[2]{#2}
\newcommand {\omi}{}

\begin{document}
\title{Very Light Jets \\ I. Axisymmetric Parameter Study and Analytic Approximation}
\author{M.Krause}
\institute{Landessternwarte K\"onigstuhl, D-69117 Heidelberg, 
  Germany}
\offprints{M.Krause,
\email{M.Krause@lsw.uni-heidelberg.de}}
\date{Received \today / Accepted <date>}
\abstract{
The propagation of extragalactic jets is studied by a series of twelve 
axisymmetric hydrodynamic simulations. 
\amd{Motivated} by observational constraints, but unlike most previous simulations,
the regime of jet to external medium density ($\eta$) from $10^{-5}$ to $10^{-2}$
is explored, for Mach numbers ($M$) between 2.6 and 26.
The computational domain contained the bow shocks for the whole simulation time.
The bow shocks are found to be spherical at source sizes below a critical value
$r_1$ (blastwave phase), which can reach up to 10 jet radii. 
After that, their aspect ratio rises slowly, as long as the bow shock stays supersonic.
The cocoons expand typically to almost the same size as the bow shock,
unless the Mach number is below \amd{approximately three}. 
Low values for the aspect ratio and the cocoon--to--bow--shock width ratio
is demanded by recent Chandra X-ray observations
of the bow shock in the archetypical radio galaxy Cygnus~A.
\cl{Therefore, $\eta<10^{-3}$ and $M<6$, in this source.}
The numerical work is complemented by an analytic approach for the spherical
phase. 
Extending previous work, the radial force balance could be integrated for arbitrary background
density and energy input, which results in a global solution. 
The analytic results are shown to be consistent with the numerical work,
and a lower limit to $r_1$ can be \amd{calculated}, which falls below the numerical
results by a few jet radii. It is shown explicitely how a King density distribution
changes the discussed aspects of the bow shock propagation.
\genco{2}{Because the jet head propagates very fast in the blastwave phase, it turns out that it
is not possible to ``frustrate'' a jet by a high density environment. 
This is very important for the class of small radio galaxies (compact symmetric objects /
GHz peaked sources): They have to be young.
During its blastwave phase, a powerful jet can transfer typically $10^{60}$~erg to
the environmental gas. This is enough to balance the radiative losses in a cooling flow,
if one of the cluster galaxies harbours a powerful jet every $10^9$ years.}  
 }

\authorrunning{M. Krause}
\maketitle
\section{Introduction}
Extragalactic jets have been known in the optical since the famous 
observation of the jet beam in M~87
\citep{Curtis1918}. The understanding of those collimated flows 
improved considerably with the advent of radio astronomy \citep[e.g. ][]{BM54}.
During that era, plenty of radio jets have been found in extragalactic objects.
Surrounding the beams, a large cocoon with lower surface brightness was found, which
can extend up to tens of jet radii \citep[e.g. in Cygnus~A][]{CarBar96}.
They have been imaged down to the very center of the mainly elliptical host galaxies,
where a supermassive black hole is believed to power the bipolar, relativistic, 
outflows 
by an accretion process. Jets can propagate up to several Mpc \citep{Schoenea01}.
According to standard jet theory, they displace the medium \omi they 
propagate \amd{into} and form a bow shock around the system.
Although, the presence of bow shocks could be anticipated from some ROSAT images
\citep{CPH94},
the Chandra X-ray observatory provided the first clear view on those features 
with \omi satisfactory resolution \citep{Sea01}. 
The bow shock in Cygnus~A is a very fine example. The source is believed to be located
nearly perpendicular to the line of sight \citep{Krea98}. 
Therefore, one can easily measure the true \omi aspect ratio (bow shock diameter 
in jet direction over perpendicular  bow shock diameter) from the Chandra image
\citep{Sea01}, which turns out to be 1.2. 
Another parameter that can be easily measured with the new Chandra data
is the relative extention of radio cocoon and bow shock. Defining this ratio
to be taken perpendicular to the jet axis through the core renders it independent
from the viewing angle.
One can in principle relate these observables to jet parameters via numerical
simulation results. However, this is only possible if the bow shock stays within
the computational domain during the simulation. Up to now only a few hydrodynamical 
simulations in the literature meet this requirement. Two examples can be found in 
\citet{RHB02} and \citet{mypap02a}. In the latter case, a jet with a density 
contrast (jet over surrounding medium) $\eta=10^{-4}$ was simulated. This resulted in an aspect
ratio of slightly more than unity, similar to Cygnus~A.  

Jet simulations have been performed so far mainly with $10^{-2}<\eta<10$
\citep[e.g. ][]{Normea82,Normea83,CNB86,KM88,Lindea89,CB92,Clarke93,Masea96,Kom99,TJR01,mypap01a}
\gc{.}
Those studies have been reviewed by \citet{Norm93} and \citet{Fer98}.
One main result was that light jets, \cl{i.e.} less dense than the surrounding medium, develop
extensive cocoons, and are therefore believed to be present in radio galaxies and quasars.
The regime below that density, which will be called {\em very light}, was touched only 
occasionally. 

\genco{1}{Recently, parameter studies of jets were carried out by \citet{CarvOdea02a,CarvOdea02b}
and \citet{Saxea02}. These studies partly employ an open boundary condition on the 
jet injection side. In that respect, they are complementary to the study presented
here. The problem with the open boundary condition is that the backflow can transport 
large amounts of energy off the grid. This is a problem in the case of very light jets
(especially for young bipolar sources),
since the backflow is very fast compared to the head propagation.
A similar problem arises when the bow shock leaves the grid sideways.
\citet{Saxea02} have presented simulations with a closed injection side boundary
and a big enough grid for the bow shock to remain inside for most of the simulation time.}

In this paper a systematic study of very light jets with 
$\eta \in [10^{-5};10^{-2}]$ and internal Mach number 
$M_\mathrm{j} \in [2.6; 26]$
is presented. 
\genco{1}{The injection side boundary is closed, and the grid is big enough to follow
the bow shock for the whole simulation time.
Contrary to previous studies, the focus in this paper is on the jet physics 
specific for very low jet densities.}
First, observational constraints on the jet parameters 
are reviewed. It follows a description of the simulations.
Afterwards, an analytic approximation for the bow shock propagation is given,
and finally the results are discussed.
The time unit used in the following is the Myr, an abbreviation for $10^6$ years.

\section{Constraints from Observation}
\label{obconst}
In nearby radio galaxies, the movement of individual jet features can be observed
in the inner \amd{100} parsecs \citep[e.g.][]{Britzea00}.
These measurements indicate apparent superluminal motion, which is a unique feature
of relativistic velocities, very close to the speed of light $c$.
The velocity at kpc scale is not so easy measurable. \citet{CarBar96} cite a
value of $v_\mathrm{J} \approx 0.4 \,c$ for the jet velocity in the kpc scale jet
of Cygnus A, estimated from the hot spot spectra and consistent with luminosity,
minimum energy hotspot pressure, lack of internal Faraday dispersion in the lobes,
and the jet-to-counterjet surface brightness ratio. However, it should be emphasised,
that this number is quite uncertain.

\citet{Parmea99} measure the radio source lifetime by synchrotron aging.
Their straight forward analysis yields a lifetime $t_\mathrm{s}<100\, \mathrm{Myr}$ for low power
sources and  $t_\mathrm{s}<10\, \mathrm{Myr}$ for high power ones. However, they point out
that this discrepancy does not have to be real, since the estimate depends on the
magnetic field, which is more uncertain in high power radio galaxies.
Reducing the magnetic field to one fourth of the applied equipartition value
\amd{pushes} the high power sources to 100 Myr, also. This would be supported
by jet-to-counterjet length asymmetry constraints \citep{Scheuer95},
and just be consistent with the age of the universe in the case of extended sources
at the highest redshifts.
Since there is the possibility that the older populations of synchrotron electrons
are polluted by reaccelerated ones or mixed with younger populations in backflow
regions, synchrotron ages indicate lower limits.
Adopting 100 Myr as fiducial upper limit for radio galaxy lifetimes,
\cite{Parmea99} arrive at a typical head advance speed of
$v_\mathrm{head}=(0.5-5.0)$ kpc/Myr for the low power sources.
It is not unreasonable that the same parameters apply for high power ones.
For those, \citet{Scheuer95} gives an upper limit of $v_\mathrm{head}=30$ kpc/Myr,
from jet-to-counterjet length asymmetry measurements.

The advance speed of the bow shock in jet direction can be computed from the one\amd{-}dimensional
momentum balance:
\begin{equation}
\label{v.head_exact}
\frac{v_\mathrm{j}}{v_\mathrm{b}} = \frac{1}{\sqrt{\eta \epsilon}}
\sqrt{\frac{1+1/\gamma_\mathrm{m}
M_\mathrm{b}}{1+1/\gamma_\mathrm{j} M_\mathrm{j}}} +1.
\end{equation}
Here, $\epsilon$ is the ratio of beam to head cross section
(simulations determine this parameter to $\epsilon \in[0.1;1]$
for $\eta>0.01$ \citep[e.g. ][]{Normea83}), $v_\mathrm{b}$ and $M_\mathrm{b}$ 
are velocity and Mach number (with respect to external medium) of the bow shock.
$\epsilon$ is a measure for the propagation efficiency.
$\gamma_\mathrm{m}$ and $\gamma_\mathrm{j}$ are the adiabatic indices of external medium 
and jet beam, respectively.
Equation (\ref{v.head_exact}) can be approximated for high Mach numbers and low jet density by
\begin{equation}\label{v.head_approx}
v_\mathrm{head}=\sqrt{\eta\epsilon} \,v_\mathrm{j} 
\end{equation}
$\eta$ is constrained by the fact that radio lobes develop 
and the above considerations 
\speco{1}{(Eq. (\ref{v.head_approx})):
$3 \times 10^{-6} < \eta < 1$.
However, it will be shown below that Eq. (\ref{v.head_approx}) is
no longer valid at these low jet densities.
A lower limit of $10^{-7}$ for the density contrast is derived in Sect.~\ref{disc}.}
Unfortunately, the temperature in the jet beams is not known. Therefore, 
Mach numbers are unconstrained, observationally.

Jet radii ($r_\mathrm{j}$) are of the order of kpc \citep[e.g.][]{CarBar96,Scheckea02}.
From the total radio luminosities, one can estimate the total kinetic jet luminosity.
Since the source needs at least the energy to inflate the radio lobes,
the total power has to be at least about twice the radio power
\citep[for Cygnus A,][]{CarBar96}, probably more.
The most powerful sources at high redshift should therefore have $L_\mathrm{kin}
>10^{46}$ erg/sec.

\section{Simulation Setup}
For an investigation of the parameter space of very light jets, axisymmetric
hydrodynamic simulations were performed.
The Newtonian 3D MHD code NIRVANA \citep{ZY97} was used for the computations.
NIRVANA is second order accurate and \amd{employs}
a monotonic upwind differencing scheme.
It treats
the following standard set of hydrodynamic equations:
\begin{eqnarray}
\frac{\partial \rho}{\partial t} + \nabla \cdot \left( \rho {\bf v}\right)&
 = & 0 \\
\frac{\partial \rho {\bf v}}{\partial t} + \nabla \cdot\left( \rho {\bf v}
{\bf v} \right) & = &
- \nabla p \\
\frac{\partial e}{\partial t} + \nabla \cdot \left(e {\bf v} \right) & = &
- p \; \nabla \cdot {\bf v}, \label{ie}
\end{eqnarray}
where $\rho$ denotes the density, $e$ internal energy density,
${\bf v}$ velocity, and $p=(\gamma -1) e$ the pressure. Here
$\gamma = 5/3$ for a nonrelativistic gas is assumed.

For a first scan of the parameter space 12 axisymmetric simulations
were performed. The jets were injected in pressure equilibrium into a homogeneous external
medium. 
\begin{table}[h]
\caption{Simulation parameters}
\label{simparams}
\vspace{-3mm}
\begin{center}
\begin{tabular}{cccc}  \hline \hline
\multicolumn{4}{c}{\vspace{-3mm}}\\
$\eta$    & $v_\mathrm{j}$ [kpc/Myr] & $M_\mathrm{int}$ & $M_\mathrm{ext}$ \\ \hline
\multicolumn{4}{c}{\vspace{-3mm}}\\
$10^{-2}$ & 95                     &  25.56	       &  255.6  \\
$10^{-2}$ & 31                    &   8.083	       &   80.83 \\
$10^{-2}$ & 9.5                    &   2.556	       &   25.56 \\ \hline
\multicolumn{4}{c}{\vspace{-3mm}}\\
$10^{-3}$ & 95                     &  25,56 	       &  808.3  \\
$10^{-3}$ & 31                     &   8.083	       &  255.6  \\
$10^{-3}$ & 9.5                    &   2.556	       &   80.83 \\
\hline
\multicolumn{4}{c}{\vspace{-3mm}}\\
$10^{-4}$ & 95                      & 25.56	       & 2556    \\
$10^{-4}$ & 31                     &  8.083 	       &  808.3  \\
$10^{-4}$ & 9.5                     &  2.556	       &  255.6  \\
\hline
\multicolumn{4}{c}{\vspace{-3mm}}\\
$10^{-5}$ & 95                      & 25.56	       & 8083    \\
$10^{-5}$ & 31                      &  8.083	       & 2556    \\
$10^{-5}$ & 9.5                     &  2.556	       &  808.3  \\
\hline
 \hline
\end{tabular}
\end{center}
\end{table}

The hydrodynamic simulations are fully determined by the internal Mach number $M_\mathrm{int}$,
the density ratio $\eta$, and the pressure ratio, which is unity.
Hence, they are scalable to the parameter range needed in a specific source.
Table~\ref{simparams} gives Mach number ($M$), density contrast ($\eta$), and
the applied  velocity. The density in the external medium was set to 2~cm$^{-3}$,
and the jet radius to
$r_\mathrm{j}=1\,\mathrm{kpc}$\footnote{
The scaling was chosen in order to be easily comparable to previous work, which was adapted 
to parameters assumed to be present at high redshift \citep{mypap02a}.}. 
The grid of $[600\times600]$ points covered an area
of $Z \times R= 30\times 30\, \mathrm{kpc}^2$, so the jet radius corresponds
to 20 grid points.
Boundary conditions were set to axial symmetry on the axis,
reflecting on the left-hand side, and open on the other sides.
The reflecting boundary on the left-hand side is essential. If there was an open
boundary, a substantial amount of energy would leave the grid on that side.
The simulation was stopped when one of the following events happened:
\begin{enumerate}
\item 	The jet propagated to the right-hand side of the grid.
\item 	The computation time exceeded a reasonable amount without
	promising new behaviour in the near future.
	Typically, this time was several weeks.
\item	The jet stopped at the nozzle, producing a shock at the inlet which ignored 
	the shock jump conditions. This was caused by entrained material
	from the shocked external medium approaching the nozzle.
\end{enumerate}
\begin{figure*}[t!]
\centering
 \includegraphics[width=17cm]{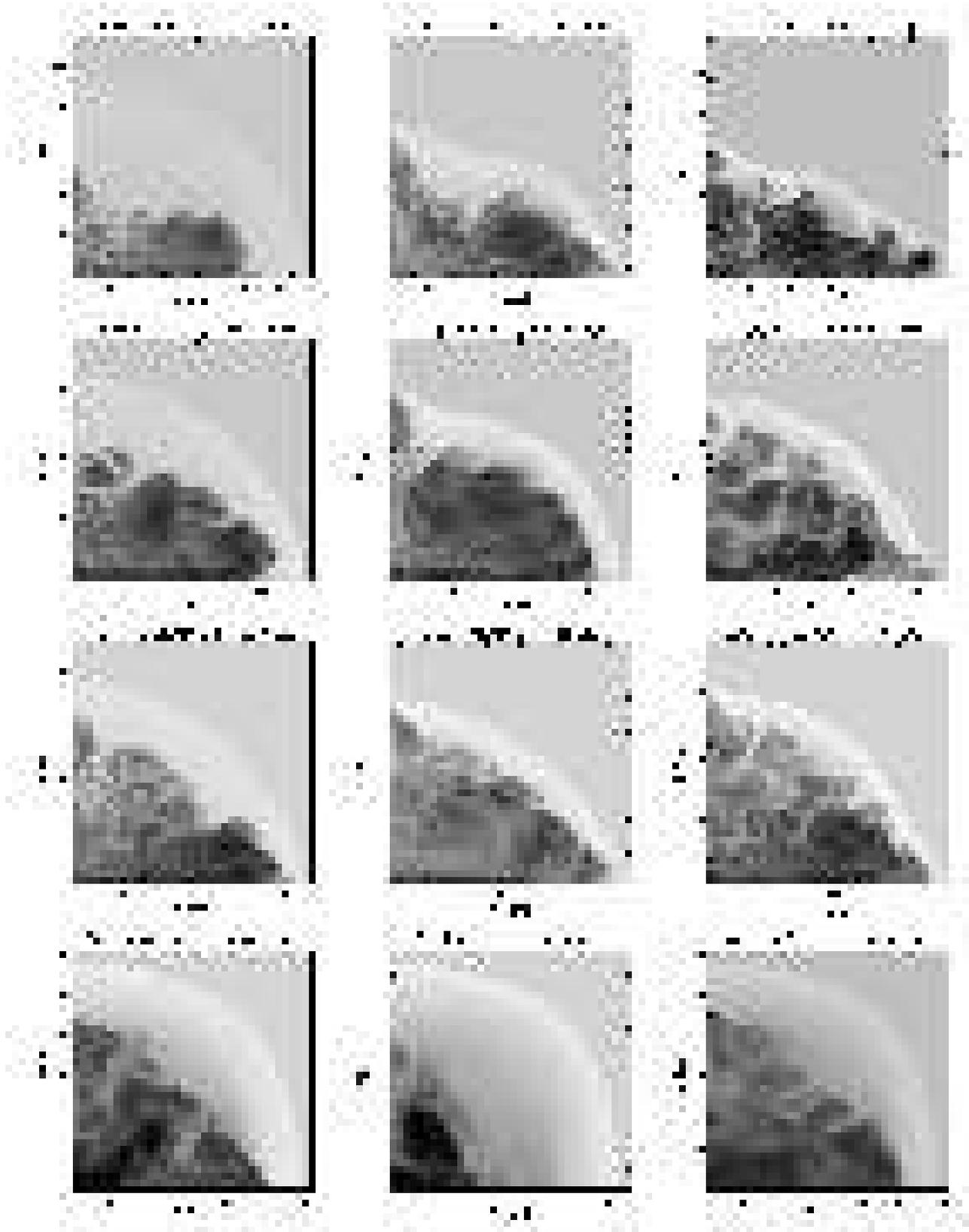}
 \caption{\small 
	Overview over the 12 simulations performed in order to scan the parameter space.
	The Mach number ($M$) varies from 2.6 to 26 from left to right.
	Density contrast ranges from $10^{-2}$ (top) to $10^{-5}$ (bottom). The logarithmic
	density contours vary from dark (low) to white (high).
	The white spot in the center of the big vortex in 
	the $(M,\eta)=(8,10^{-5})$  plot as well as several 
	jet inlets are artificially white because of a problem in the plotting routine.
	In fact, they have the lowest values. The simulation time is indicated on top of the
	individual plots.}
  \label{densheet}
\end{figure*}
\section{Results}
\subsection{Cocoons}
Contour plots of the final timestep for each simulation are shown in Fig.~\ref{densheet}.
As expected from higher $\eta$ simulations, the cocoons broaden with lower $\eta$.
\citet{BC89} derive an analytic formula for the cocoon width $r_\mathrm{coc}$:
\begin{equation}\label{cocwid}
r_\mathrm{coc}/r_\mathrm{j} \approx 0.7 M_\mathrm{j} \eta^{-1/4}.
\end{equation}
Figure~\ref{densheet} shows that only the \amd{$(M,\eta)=(2.6,10^{-2})$} cocoon can be described 
properly by the above formula. For the other jets, the bow shocks are still too small
to contain cocoons of the predicted size.
Also, the geometry of the cocoon is quite different from the cylindrical one
known from higher $\eta$ jets.
The density distributions show more or less spherical cocoons. \speco{3}{} \omi
Between $\eta=10^{-2}$ and $\eta=10^{-3}$, the cocoon undergoes a transition.
Sometimes, the vortices it dissolves in are stringed in a line around the beam, but sometimes
they join together forming a big vortex which extends approximately over the 
same size as the jet.
Figure~\ref{densheet} 
$(M,\eta)=(2.6,10^{-3})$ shows a new vortex just before
being swallowed by the big one. A vortex with substructure is present in
$(M,\eta)=(8,10^{-3})$.
The interface between cocoon and shocked external medium suffers from 
\amd{Kelvin-Helmholtz} (KH) instabilities. 
They start at small size in the vicinity
of the jet head. As they grow, extending their fingers into the cocoon, the backflow 
accelerates them towards the center. 
This behaviour is known from higher $\eta$ simulations \citep{mypap01a} at high resolution.
On the left-hand boundary, the fingers can get long enough
to interact with the beam directly. In some cases, this caused a strong shock 
at the jet inflow,
efficiently pushing the jet out of the computational domain
(one of the reasons, why the simulation had to be stopped, see above).
\speco{3 \& 4}{Quite often, this gas is drawn towards the jet 
head in between cocoon and jet beam, thus separating the cocoon from the beam.}
In the $\eta=10^{-5}$ simulations, the initial conditions are not yet relaxed enough
to determine the cocoon morphology reliably.

\begin{figure*}[t!]
\centering
\begin{minipage}{\textwidth}
 \rotatebox{-90}{\includegraphics[width=.35\textwidth]{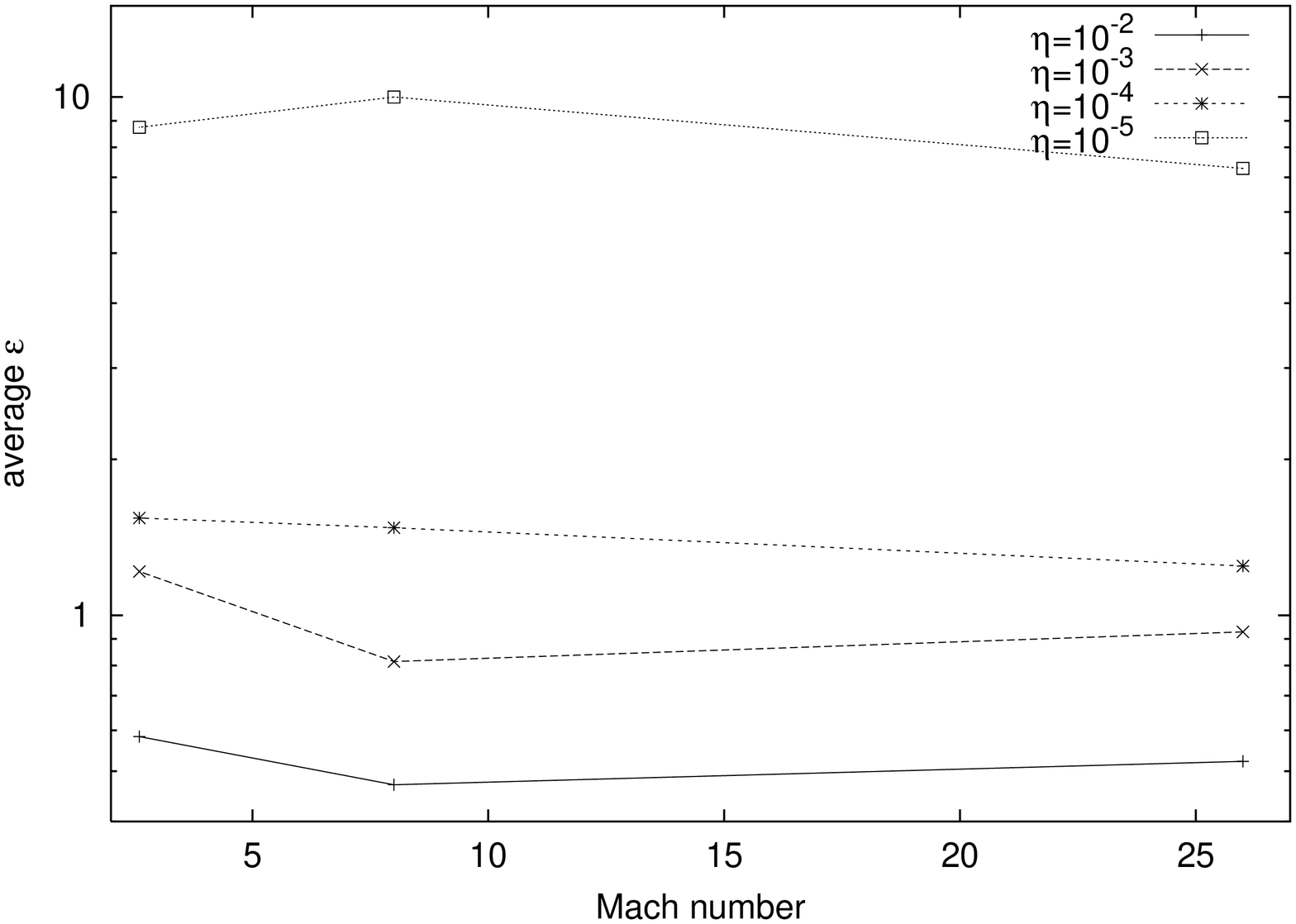}}
 \rotatebox{-90}{\includegraphics[width=.35\textwidth]{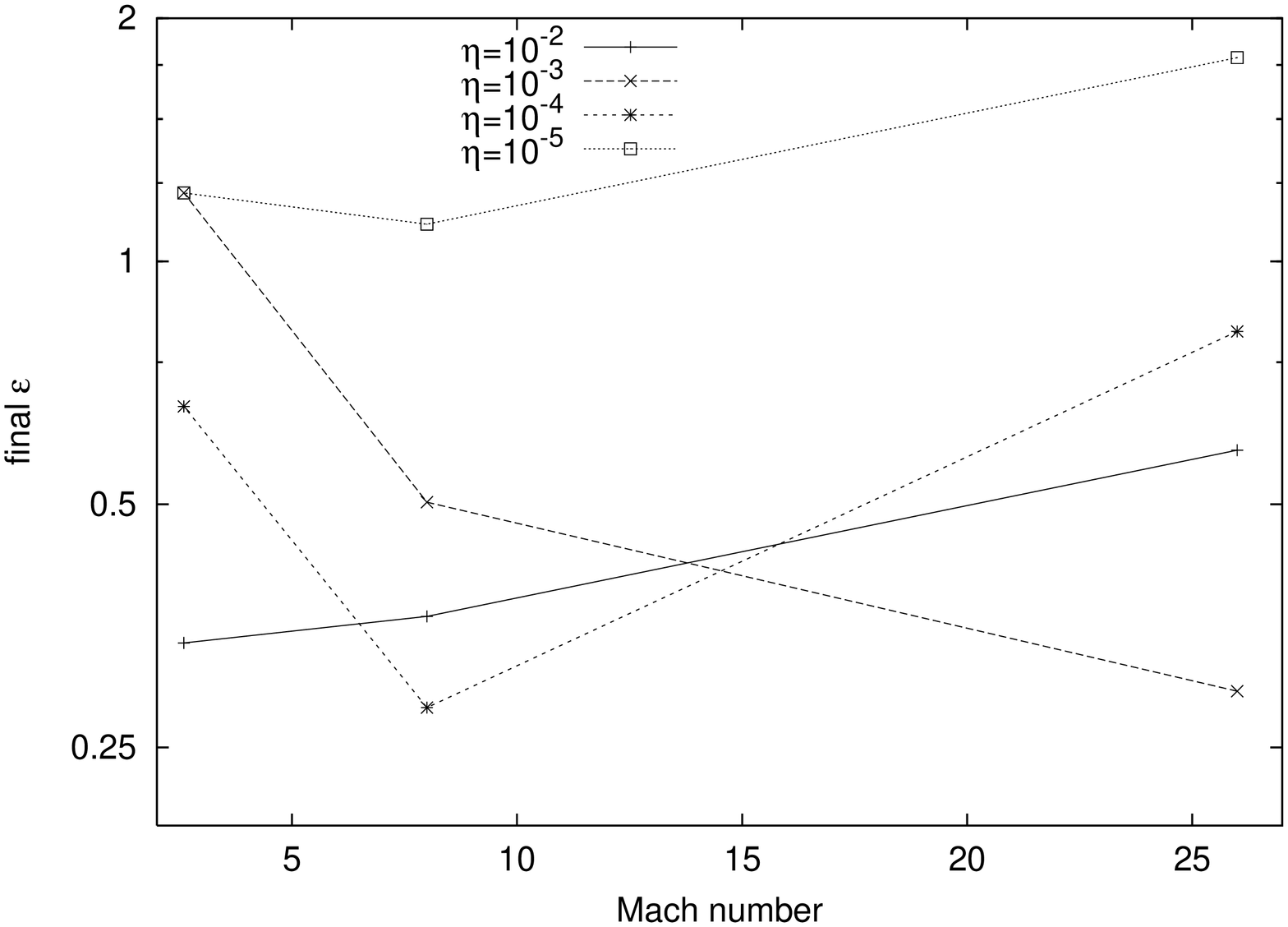}}
 \caption{\label{prop_effies}\small The propagation efficiency measure $\epsilon$ 
	on average (left panel) and for the
	last 5\% of the simulation time (right panel).}
\end{minipage}
\end{figure*}

\subsection{Bow shocks}
The bow shocks in the presented simulations are quite different from one another and
from simulations with higher $\eta$. Morphologically, they can be classified in three
groups. The extremes are located in the upper corners and the bottom row of Fig.~\ref{densheet}.
The jet in the upper right corner $(M,\eta)=(26,10^{-2})$
shows an elongated bow shock, comparable to
the simulation of \citet{Lokea92}. Due to the high Mach number, the compression ratio 
is four over the whole surface of the bow shock.
Reducing the Mach number to $(M,\eta)=(2.6,10^{-2})$, 
the bow shock gets very regular (upper left corner).
Evidently, this is caused by the lag of the contact discontinuity behind the bow shock.
Even though there are some disturbances in the shocked external medium,
they are not fast enough to catch the leading edge. 
The bow shock is still stronger in the direction of the jet propagation which produces
a higher compression ratio near the head.

\begin{figure*}[t!]
\centering
 \includegraphics[width=18cm]{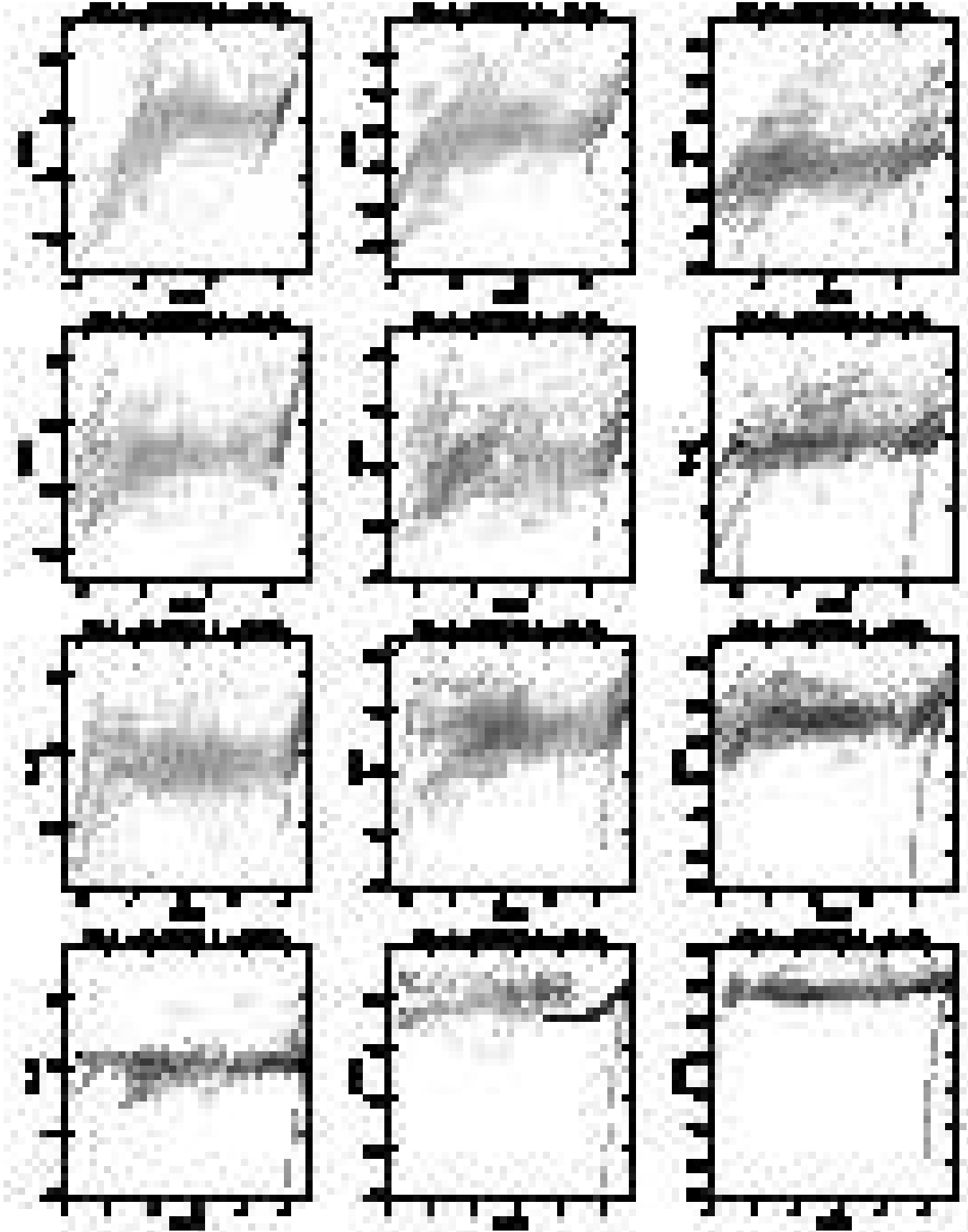}
 \caption{\small  Pressure over number density (pn) histograms.
	100 bins were used on each axis. Pressure, number density, and 
	counts are given in logarithmic units. Darker regions have more counts.
	\speco{12}{Pressure and density have been normalised to the initial jet values.}}
  \label{pn_histos}
\end{figure*}

If one would reduce the Mach number below one, no jet would form at all, and a spherical
sound wave would propagate outwards.
\speco{6}{
In the bottom row ($\eta=10^{-5}$), the bow shock is spherical, reminding one of
a spherical wave. Nethertheless, one can clearly tell from the density jumps
that it is truly a bow shock.
So, why is it spherical? }
A hint comes from the pressure versus number density (pn) histogram
(Fig.~\ref{pn_histos}). From their start position at the external pressure 
and the density of the jet and external medium ($10^5$ times jet density), respectively
the fluid elements move upwards to a thin line of high constant pressure.
(For the $(M,\eta)=(2.6,10^{-5})$ and $(M,\eta)=(8,10^{-5})$ plots, the pressure 
increases so rapidly in the jet that the counts are too few to show up in the plots.)
The pressure equilibrium of all the fluid parts affected by the jet can be understood,
if one considers that the sound speed in the jet (\amd{roughly} $v_\mathrm{jet}/2$) is much
higher than the 
\speco{7}{bow shock propagation speed. 
Therefore, sound waves communicate and equalise pressure differences efficiently.}

The internal energy within the bubble of the bow shock
is sufficient to drive a blastwave. The solution for the blastwave is derived in
Sect. \ref{blastwave}. In this limit the jet propagates slowly, and mainly transforms
its constantly provided kinetic energy into heat. The resulting pressure drives
the spherical blastwave. The velocity of the blastwave is proportional to $t^{-2/5}$.
Hence, the blastwave is faster than the jet bow shock at early times, and slower later on.
Equation (\ref{outbreak}) determines the minimum bow shock radius at which deviation
from the spherical blastwave can be expected. For $\eta=10^{-5}$ it turns out to be 
$9.25\, r_\mathrm{j}$.
The $\eta=10^{-5}$ jets are already slightly above that limit.
But the jets are in a quite complicated
phase at the time shown in Fig.~\ref{densheet}. 
For example in the case of $(M,\eta)=(26,10^{-5})$,
one shock in the middle of the jet just
became so strong that it developed a backflow there. The old jet head at $Z\approx 9$ kpc
is dissolving. This is a short phase, and because the blastwave swept much of the
matter away, the new jet will soon reach the position of the old one.
Then, it will presumably influence the bow shock. That this happens soon after 
the minimum radius of influence ($r_1$) is reached, 
is evident from the $\eta=10^{-4}$ density plots,
where $r_1=5\, r_\mathrm{j}$, and from the aspect ratios (see below).
One can understand the bow shock shapes of the other simulations as combinations
of these three types.

The propagation efficiency measure $\epsilon$ was determined from the simulations 
on average and for the final 5~\% of simulation time. The result is shown in Fig.~\ref{prop_effies}.
The highest values of $\epsilon$ are reached for $\eta=10^{-5}$.
In that case the bow shock is on average three times faster than the maximum
expected from the one-dimensional estimate.
Again, this shows the high velocity of the blastwave.
Also for the $\eta=10^{-4}$ and the $\eta=10^{-3}$ 
simulations the average efficiency is still high, \amd{for the same reason}.
The $\eta=10^{-2}$ simulations have already reached an extension of \amd{roughly} $15$
times their minimum interaction distance. Therefore $\epsilon$ has fallen
considerably below one. Also, it does not change significantly for the final 5~\%
of the simulation time. Slight variations in the propagation speed are typical for
such jets. Extrapolating from higher $\eta$, 
the accuracy of $\epsilon$ is a few percent for the given resolution
\citep{mypap01a}.
\omi
Summarising, very light jets blow up a big bubble, rapidly. After reaching $r_1$,
they stroll along with nonspectacular $\epsilon$.

\begin{figure}[b!]
\centering
\resizebox{\hsize}{!}{
 \rotatebox{-90}{\includegraphics[height=7cm]{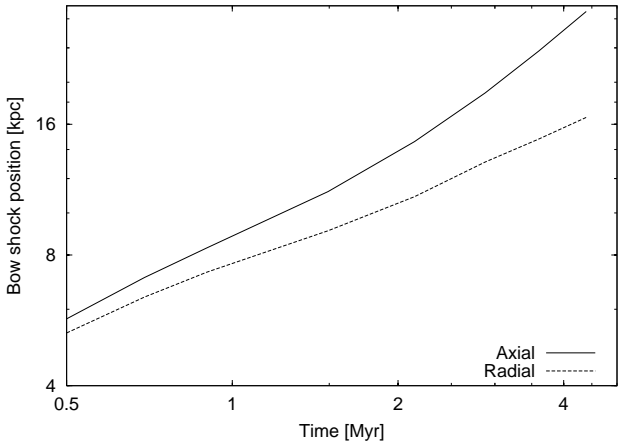}}}\\
 \caption{\label{bopro}\small  Bow shock propagation over time for the $(M,\eta)=(26,10^{-2})$ simulation.
	Shown is the position reached on the R and Z-axis, respectively.
	Logarithmic units were used for both axes.}
\end{figure}

\begin{figure}[bht]
\centering
\resizebox{\hsize}{!}{
 \rotatebox{-90}{\includegraphics[height=7cm]{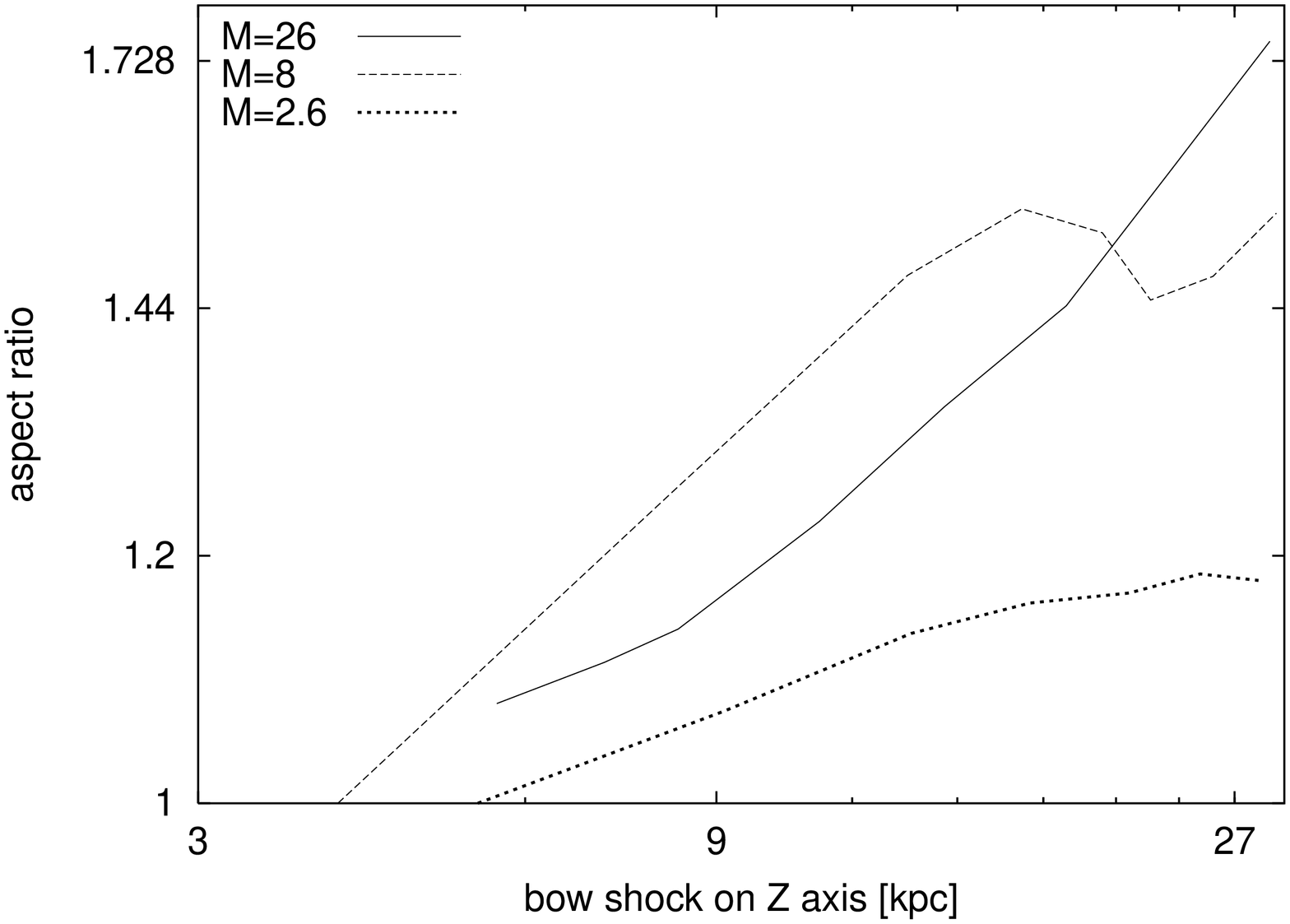}}}\\
\resizebox{\hsize}{!}{
 \rotatebox{-90}{\includegraphics[height=7cm]{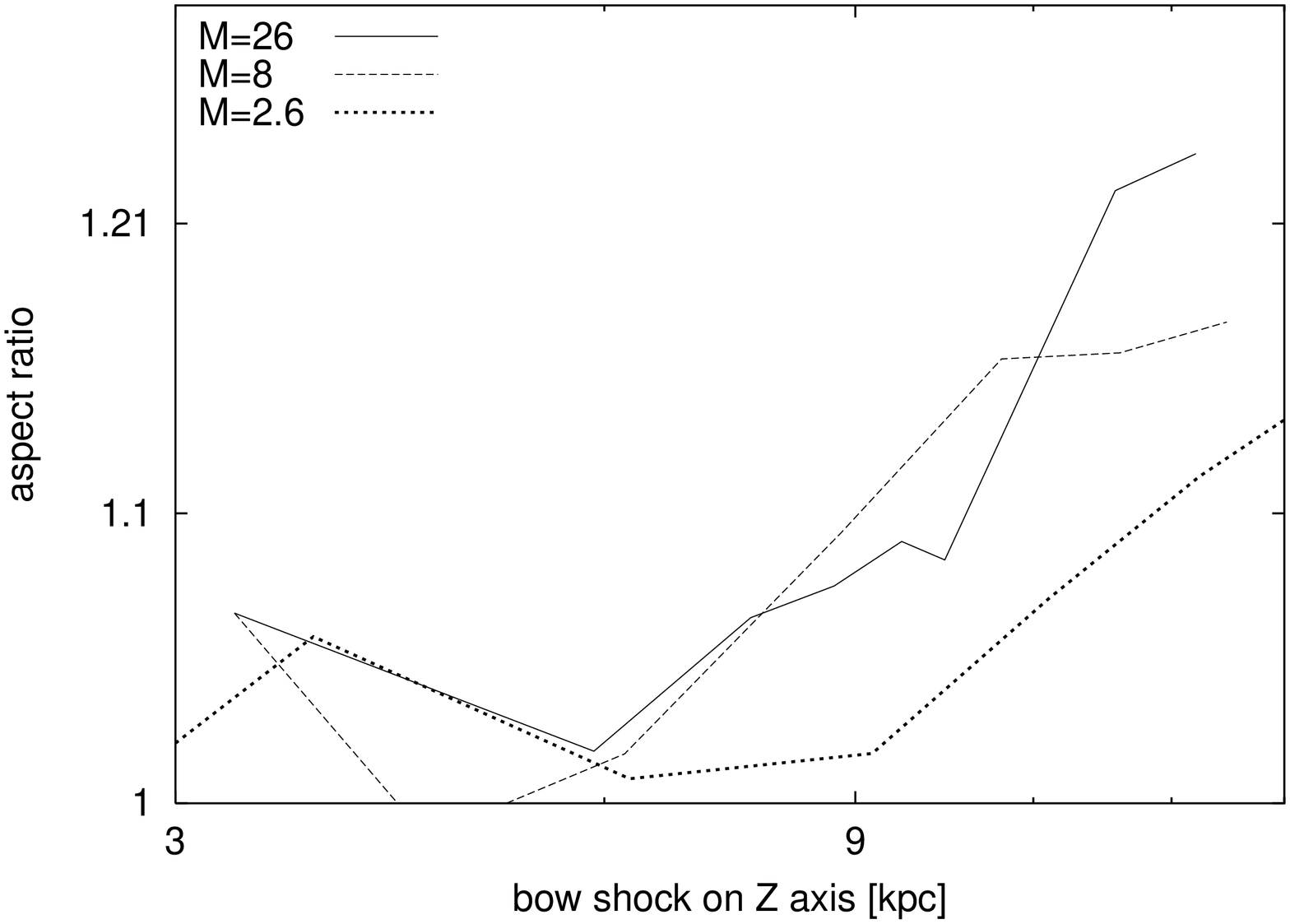}}}\\
\resizebox{\hsize}{!}{
 \rotatebox{-90}{\includegraphics[height=7cm]{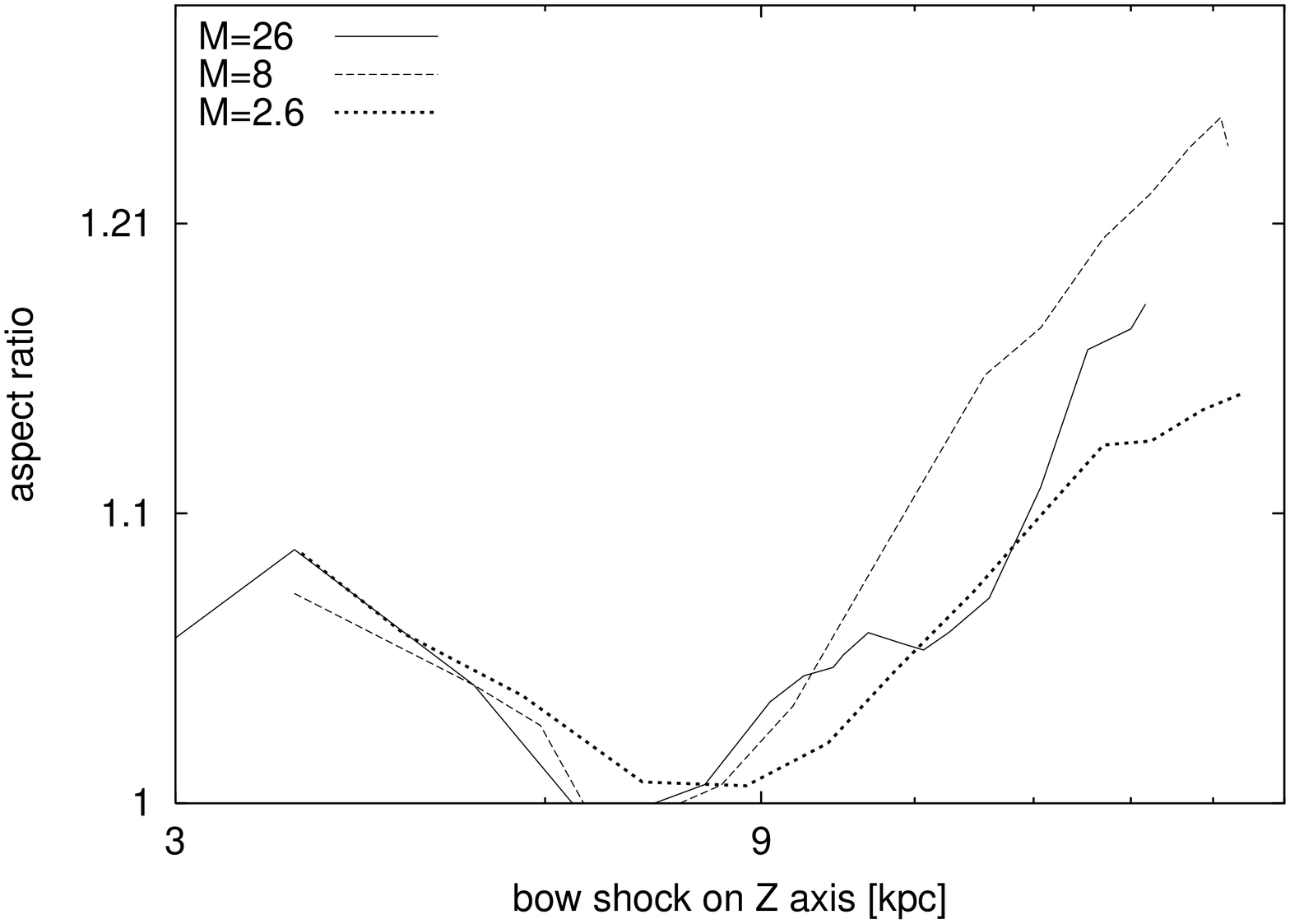}}}
 \caption{\label{aspect}\small  Aspect ratio of the bow shock over propagation distance. 
	The upper panel shows the $\eta=10^{-2}$ simulations, the middle one the
	$\eta=10^{-3}$ simulations, and the $\eta=10^{-4}$ simulations are shown in the bottom panel.
	Logarithmic units were used for both axes.}
\end{figure}

The aspect ratio of the bow shock was measured in regular intervals,
and is shown in Fig.~\ref{aspect} 
as a function of the position of the bow shock on the Z-axis. 
A general trend is evident for all 
simulations:  Up to a propagation distance
of two to three jet radii above $r_1$ the aspect is approximately one. 
This is also consistent with 
the $\eta=10^{-5}$ simulations. Then the aspect starts to rise. 
This can be explained by considering the bow shock expansion 
in the axial and radial directions. Up to $r_1$, the bow shock has to expand 
according to the blastwave law, in all directions. After that phase,
it changes to a typical jet propagation law, for the axial direction
(compare Fig.~\ref{bopro}).
The jet head sometimes accelerates temporarily,
which is well-known in jet physics ({\em beam pumping}, compare e.g.
\citet{KM88}), but has on average a constant velocity, given by 
Eq.~(\ref{v.head_approx}).

\speco{8}{
Concerning the radial direction, the bow shock is still pressure driven. 
It will be shown in Sect.~\ref{blastwave} (Eq. (\ref{mbow}))
that the Mach number of the bow shock at
$r_1$ with respect to the external medium is given by the jet's Mach number $M_\mathrm{int}$.
Hence, the bow shock is still strong at $r=r_1$, at least for the $M>3$ simulations.
The deceleration is proportional to the aspect ratio via the pressure
(assuming elliptical deformation), and should therefore be higher.
On the other hand, secondary shocks in the shocked ambient gas can now 
reach the surface of the bow shock, accelerating it in limited regions.
As an example, 
for the ($M,\eta$)=($8,10^{-4}$) simulation, the relative decrease of the radial bow shock
velocity ($1-v_\mathrm{b}/v_\mathrm{a}$)
between radial bow shock positions $r_a=6.65$~kpc and $r_b=17.0$~kpc amounts to 47.8~\%.
One can easily derive from Eq.~(\ref{vofr}) that the blastwave law predicts
a decrease of only $1-(r_\mathrm{a}/r_\mathrm{b})^{2/3}=46.5~\%$. 
The agreement between the numbers is quite good in this case.
In the other simulations there is more accumulation of jet material at the left
boundary (compare Fig.~\ref{densheet}), 
which sometimes strongly disturbs the bow shock near the left boundary.
The difference between the measured and predicted velocity decrease can reach 
\omi a factor of two \amd{there}.
However, the boundary condition at the left side complicates the interpretation.
A preliminary conclusion is therefore that the radial bow shock expansion
keeps following the blastwave law for the simulated jets with $M>3$ and $\eta< 10^{-2}$
for the simulated times, if the disturbance by the material accumulated at the left boundary is not too strong.
In a follow-up paper, where a 3D simulation with bipolar jets will be presented,
this issue will be addressed in more detail.}

Consequently, the aspects increase with time.
In Fig.~\ref{densheet}, the $\eta=10^{-3}$ and $\eta=10^{-4}$
simulations have left the blastwave phase not long ago,
whereas the $\eta=10^{-2}$ \amd{jets} \omi are already more than ten
times bigger than $r_1$. The former jets are observed to push 
comparatively \amd{gently}, basically elongating the bubble.
The $\eta=10^{-2}$ jets show an interesting dependence on the Mach number:
Only the low Mach number jet keeps the elongated spherical shape.

The cocoons of the high Mach number jets are underexpanded according
to Eq.~(\ref{cocwid}), and act violently on the bow shocks.
In the phase where the bow shock is only slightly elongated,
the bow shock velocity in axial direction 
has not yet reached the constant velocity.
This can be seen from Fig.~\ref{bopro}:
The gradual change between the two propagation laws is finished 
at $t\approx 2$ Myr, for the $(M,\eta)=(26,10^{-2})$ jet. At that 
time the bow shock has reached \amd{approximately} $10\,r_1$.
The limiting aspect ratio ($a$) can then easily be computed from Eq.~(\ref{v.head_approx}):
\begin{equation}
a=\frac{v_\mathrm{j} \sqrt{\epsilon\eta}}{c_\mathrm{s,\,ext}}
=M_\mathrm{ext}\sqrt{\epsilon\eta},
\end{equation}
where the index ext denotes quantities in the external medium.
With the values from Table~\ref{simparams} the limiting aspect ratios turn out to be
between 1.6 ($M=2.6$) and 16 ($M=26$). None of the simulations has \omi 
reached this limit \amd{yet}.

\subsection{Beams}
\speco{9}{} \omi

\begin{figure*}[t!]
\centering
 \includegraphics[width=18cm]{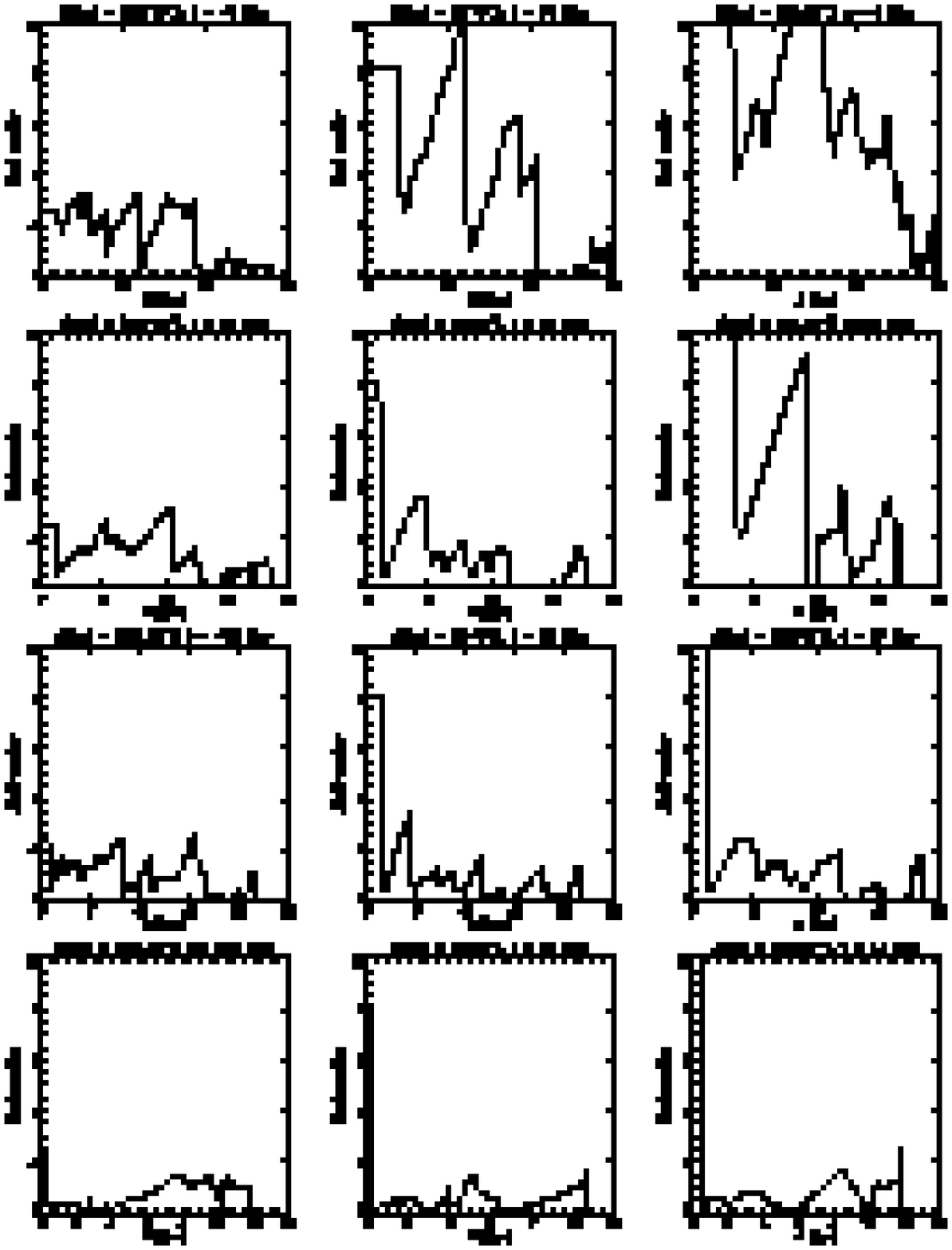}
 \caption{\small On-axis Mach number.}
  \label{mach}
\end{figure*}

\speco{10}{
A prominent feature of the jet beams is the strong shock roughly one jet radius behind
the jet inlet (compare Fig.~\ref{mach}). 
This indicates the first reflection point of the oblique shocks in the jet.
The shock is caused by the high 
pressure and velocities of the backflow, which acts on the beam. 
The position of the first shock reflection point varies with time.
This is consistent with simulation results at higher density contrast
with closed left boundary conditions \citep{KM88}.}

High Mach numbers ($M>3$) can be sustained for the $\eta=10^{-2}$ jets (compare Fig.~\ref{mach}).
With the exception of the $(M,\eta)=(26,10^{-3})$ jet, all the other beams struggle
around $M=1-2$. Quite often they become subsonic, even if they start with high 
Mach numbers. Strong shocks decelerate those jets. Since also shocked ambient gas
reaches the beam quite often,
nothing prevents the beams from being disrupted by the KH instability.
A fine example is the $(M,\eta)=(26,10^{-3})$ simulation (Fig.~\ref{densheet}).
The shocks in the beams sometimes get strong enough to shed vortices and produce backflows,
thereby establishing a new jet head. 
\speco{11}{
This behaviour is similar to axisymmetric simulations of higher $\eta$ jets
\citep[e.g. ][]{KM88,JRE99}, but seems to be more violent at lower density contrast.}

The $\eta=10^{-5}$ jets suffer from the short evolution time.
At the time shown, the jets still have the prominent shock at the jet inlet, which was
used as initial condition.
Since those are the computationally most expensive simulations,  the problem can only be solved
with a faster computer, evolving the jets longer.


\subsection{Pressure versus number density histograms}
\begin{figure*}[htb]
\centering
 \includegraphics[width=8cm]{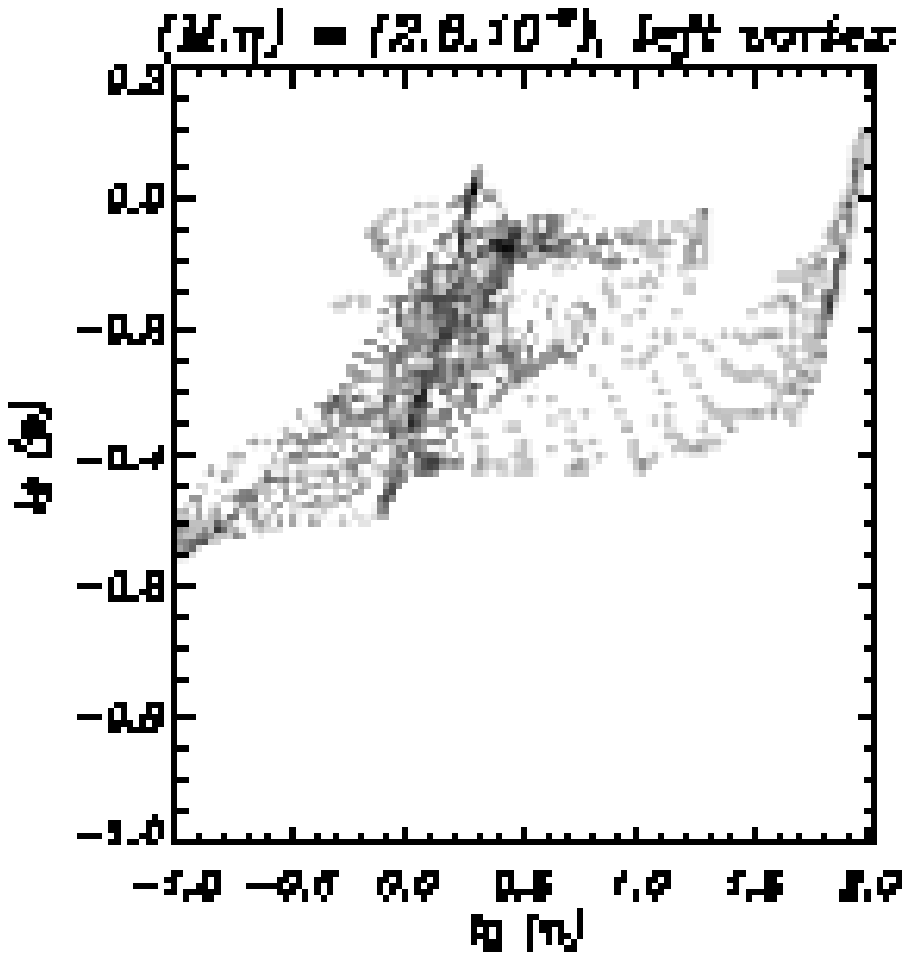}
 \includegraphics[width=8cm]{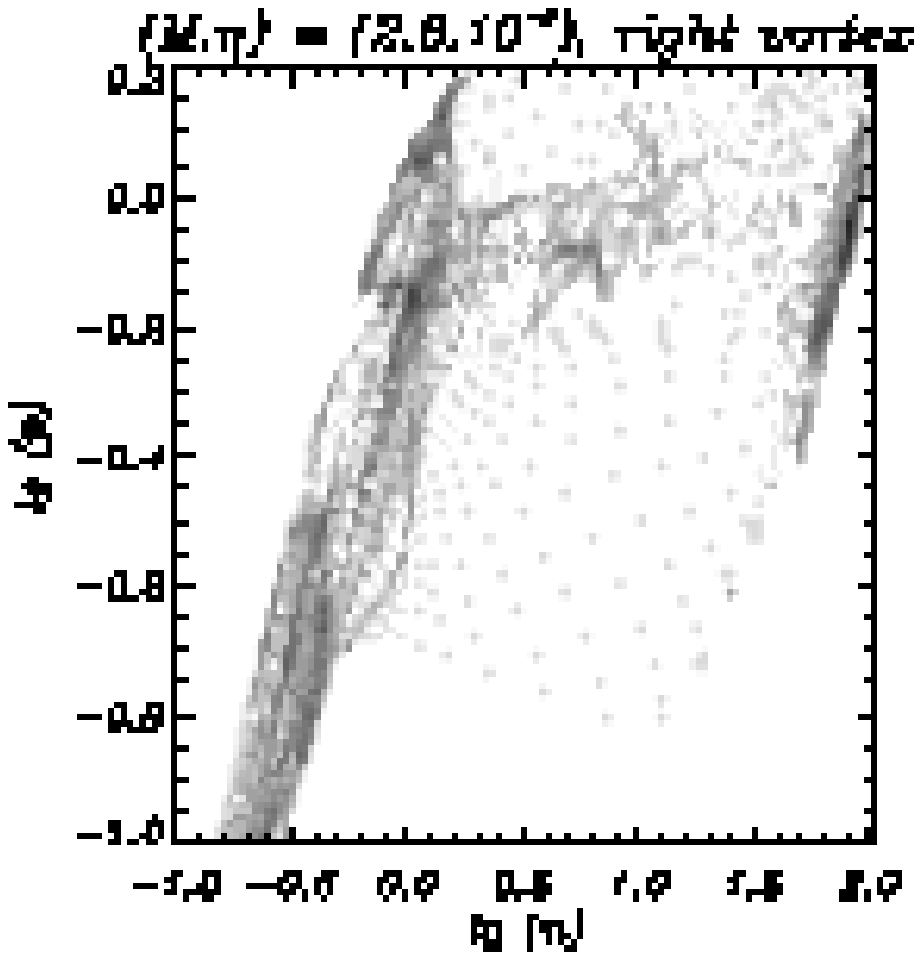}
 \caption{\small  pn-histograms for regions extracted around the two vortices
	of the $(M,\eta)=(2.6,10^{-2})$ jet. The left figure corresponds to the left vortex.}
  \label{pnsep}
\end{figure*}

From Fig.~\ref{pn_histos}, it is evident that very light jets provide their own pressure 
everywhere. External medium as well as jet plasma get strongly shocked, and the high sound speed
equalises the pressure within most of the jet bubble, especially in the higher Mach
number cases.
The external pressure is therefore unimportant, and the pressure matching,
used as initial condition, can be dropped in future simulations, without much difference.
The pn-histograms show many counts at intermediate densities. They are caused 
by fluid elements resulting from mixing of jet plasma with shocked external medium.
This is present in every simulation, although there are typically less counts than in 
the region of shocked external medium densities (right) or jet densities (left).
\speco{13}{} \omi
Especially prominent are several straight lines in the pn-histograms.
They extend to very low densities, present in the center of vortices.
The following example examines the $(M,\eta)=(2.6,10^{-2})$ jet. The density contours
show two prominent vortices (dark) in the cocoon. Correspondingly, the pn-histogram
shows two spikes extending down to a tenth of the jet density. Now, two areas, each
centered on one vortex were cut out and examined separately. 
The individual pn-histograms are shown in Fig.~\ref{pnsep}.
Clearly, in the right figure, the upper spike has disappeared. Nevertheless, it is prominent
in the left figure. Hence, it is evident that vortices show up in the pn-histograms
as straight lines. Therefore they follow the relation $p\propto n^\Gamma$.
\speco{15}{The left and right vortices of Fig.~\ref{pnsep} have $\Gamma=0.35$ and $\Gamma=1.26$,
respectively.
%
The spikes in the other histograms have a maximum $\Gamma$ of $1.5-1.7$.
If a vortex would originate from jet material with uniform density and would be compressed in
a shock with uniform compression ratio $\kappa$, it would have a uniform entropy index 
$s \equiv p/n^\gamma$. The pressure differences are caused by adiabatic changes due to the
centrifugal force. The consequence is a $p=s n^\gamma$ subsystem, which is often observed
in the pn histograms. 
The low $\Gamma$ of the two vortices in the $(M,\eta)=(2.6,10^{-2})$ is exceptional.
Systems of that kind are occasionally created in a strong shock in the vicinity of the jet axis
connected with a temporal on-axis backflow.
Such shocks are usually weaker in a 3D simulation.
The life time of these systems is short. They mix rapidly with other material
due to the limited grid resolution.}

The bow shock obeys a curved line in the pn-histogram. This is especially evident from 
the $(M,\eta)=(8,10^{-5})$ simulation. The density contours show most of the space occupied
by shocked external medium. In the pn-histogram, these fluid elements, starting at 
$(\log (n),\log (p))\approx (5,0)$, first move upwards, bending to the right.
The maximum pressure and density in the shock is reached in a rightmost pinnacle 
($(\log (n),\log (p))\approx (5.6,2.2)$).
Then pressure and density decline fast to the isobaric regime. Here the density
declines further. Even in this case,
with the smallest cocoon, 87 \% of the mass is concentrated in the region
$8.5\,\kpc<r<10.5\,\kpc$, forming a thin shell.

\section{Analytic Approximation for the Blastwave Phase}
\label{blastwave}
In the simulations \omi \amd{of} the last section the bow shock was found to resemble a spherical
blastwave up to a certain time. The simulations show that the cocoons displace most of the 
external medium into a shell. Assuming a constant velocity $v$ 
for a shell of mass {$\cal M$},
one can find an analytic approximation for the expansion
of this shell by postulating force balance at the bow shock surface:
\begin{equation}\label{forbal}
\frac{\partial}{\partial t} \left( {\cal M} v \right) = 
S\left(p_\mathrm{int}-p_\mathrm{ext}\right).
\end{equation}
Here, $S=4\pi r^2$, and $p_\mathrm{int}$ and $p_\mathrm{ext}$ denote internal and 
external pressure, respectively. The internal pressure is given by 
$p_\mathrm{int}=\frac{2}{3} (E(t)-{\cal M}v^2/2)/(4\pi r^3/3)$, where $E(t)$ is the energy
injected into the cocoon by the jet.

\genco{1}{Usually one considers self-similar solutions of the spherical force balance
\citep[e.g.][]{HRB98,Sokea02,mypap02a}.
Here, an alternative approach is presented, computing the global integral of the equation.}
 
\subsection{General solution}
In the case of negligible external pressure,
which is the case for strong bow shocks, Eq.~(\ref{forbal}) can be integrated:
\begin{equation}\label{solu}
\int_0^r {\cal M}(r) r dr = 2 \int_0^t dt_1 \int_0^{t_1} E(t_2) dt_2.
\end{equation}
Equation~(\ref{solu}) is the general solution for any sort of energy input
driving a blastwave into a medium with arbitrary density distribution, as long as both
are integrable functions. 

\subsection{Power law solutions}
A very useful type of solutions is found, if one assumes energy and density to be 
given by power laws:
\begin{eqnarray*}
E&=& {\cal L} t^d \\
\rho&=&\rho_0 (r/r_0)^\kappa
\end{eqnarray*}
The solution for the shell radius is in that case:
\begin{equation}\label{blastexp}
r=\left( \frac{(\kappa+3)(\kappa+5)}{(d+1)(d+2)} 
\frac{r_0^\kappa {\cal L}}{2\pi\rho_0}\right)^\frac{1}{\kappa+5} t^{^\frac{d+2}{\kappa+5}}
\end{equation}
For example, setting total energy and background density to a constant ($\kappa=d=0$)
gives the well-known solution for expanding supernova bubbles \citep{Sedov59}.
Jets can be assumed to deliver a constant amount of energy to their cocoons.
Therefore $d=1$ is appropriate.
With that choice, the functional dependence of the self-similar expansion law
from \citet{HRB98} is recovered. However, for density exponents between -2 and 0, the 
constant of proportionality is by a factor of 4 to 1 lower in the present solution.
This reflects the fact that it is, from the mathematical point of view, a global
solution. Therefore, for density exponents lower than -3 the bubble has to fight against
an infinite amount of mass at $r=0$. Equation~(\ref{solu}) consequently states that 
the blastwave will not expand at all. This is of course not found in a self-similar solution, 
since such a solution is always assumed to be far from the boundaries.
A relativistic gas in the cocoon, as taken into account by \citet{HRB98}, increases
the constant of proportionality by \amd{roughly} $10~\%$.
Because the blastwave is decelerating for \amd{$\kappa>-2$}, whereas the bow shock 
from a jet should propagate at approximately constant speed 
(compare Eq.~(\ref{v.head_approx})), the bow shock is first spherical,
and gets shaped by the jet only if the jet thrust can push it faster.
The expansion velocity of the blastwave $v$ from a non-relativistic  
\speco{17}{twin} jet is given by:
\begin{equation}\label{vofr}
\left(\frac{v}{v_\mathrm{j}}\right)^3=\frac{9}{4} \frac{\kappa+3}{(\kappa+5)^2} \eta_{\amd{0}}
\left(\frac{r}{r_0}\right)^{-\kappa} \left(\frac{r}{r_j}\right)^{-2},
\end{equation}
\speco{18}{where $\eta_0=\rho_\mathrm{j}/\rho_0$.}
Combining this with Eq.~(\ref{v.head_approx}), 
and choosing $r_0=r_\mathrm{j}$, results in:
\begin{equation}\label{outbreak}
\frac{r_1}{r_\mathrm{j}}= \left(\frac{16}{81} 
\frac{(\kappa+5)^4}{(\kappa+3)^2} \eta_0 \epsilon^{3}\right)^\frac{1}{\kappa-4} 
\approx 0.5 \,(\eta_0 \epsilon^3)^\frac{1}{\kappa-4}
\end{equation}
$r_1$ denotes the radius at which the jet first affects the bow shock.
The latter approximation is valid for $0 > \kappa>-2.9$
with an accuracy of 25~\%. The coefficient
goes to zero for $\kappa$ towards -3.
The minimum $r_1$ is reached for $\epsilon=1$.
It may be surprising \amd{at} the first glance that the jet head can even more easily
reach accelerating bubbles. This is due to the fact that the jet head accelerates 
always faster than the blastwave, 
as long as the assumptions leading to Eq.~(\ref{v.head_approx})
are appropriate.
It follows that spherical bow shocks are expected for very light jets, in small systems, and 
in environments where the density does not decrease faster than with $\kappa \approx -2.9$.
The latter restriction does not apply, if the density exponent turns smoothly into -3,
which is demonstrated below.

\speco{16}{The assumption of negligible external pressure holds as long as the bow shock is
much faster than the sound speed in the external medium, 
which can also be shown explicitely from the equations in this section.
The bow shock has a Mach number 
$M_\mathrm{B}$ with respect to the external medium at a radius $r$, in the 
isothermal case given by
\begin{equation}\label{mbow}
\frac{r}{r_1}=
\left(\frac{9}{4} \frac{\kappa+3}{(\kappa+5)^2} \eta_0^{-2/\kappa}\right)^{\frac{3\kappa}{\kappa^2-2\kappa-8}}
\epsilon^{\frac{3}{4-\kappa}}
\left(\frac{M_\mathrm{ext}}{M_\mathrm{B}}\right)^{\frac{3}{\kappa+2}}
.
\end{equation}
For the conditions of the simulations presented here ($\kappa=0$),
Eq. (\ref{mbow}) states that the Mach number of the bow shock exceeds the jet Mach
number when the bow shock reaches $r_1$.
If $\kappa <0$, the bow shock Mach number at $r=r_1$ is even higher for $\eta_0<10^{-2}$.
It can reach typically several times the jet Mach number. 
In an isobaric atmosphere, the relation becomes:
\begin{equation}
\frac{r}{r_1}=\left( \delta \epsilon 
\left(\frac{M_\mathrm{int}}{M_\mathrm{B}}\right)^2\right)^\frac{3}{4-\kappa},
\end{equation}
where $\delta$ is the overpressure ratio of the jet.
Again, the bow shock is supersonic at $r_1$ for supersonic jets, unless they are heavily underpressured.}

\subsection{King type solutions}
\begin{figure}[tbh]
\centering
 \rotatebox{-90}{\resizebox{!}{.9\hsize}{\includegraphics{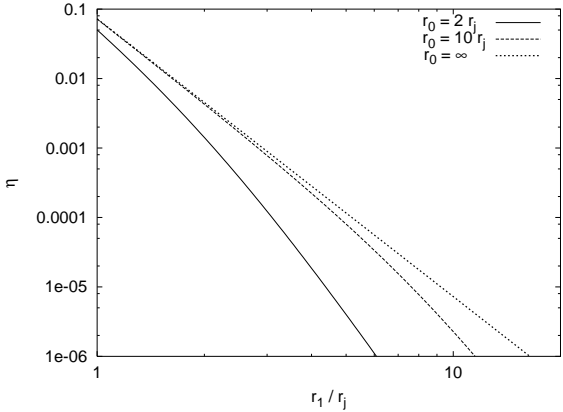}}}
 \caption{\label{r1king}\small Minimum radius of jet influence $(r_1/r_\mathrm{j})$ (horizontal axis)
	as a function of density contrast $\eta=\rho_\mathrm{j}/\rho_0$ for different core radii
	$r_0$ in a $\beta=1$ King profile. For $r_0=\infty$ the density is constant.}
\end{figure}

Extragalactic jets often \omi propagate into environments where the density cannot 
be approximated by a single power law. A King type density distribution is often used
for density \amd{distributions} around galaxies:
\begin{equation}
\rho(r)=\rho_0 \left( 1+\left(\frac{r}{r_0}\right)^2\right)^{-3\beta/2}
\end{equation}
The integrals in Eq.~(\ref{solu}) can be performed analytically 
for some cases of $\beta$, resulting in:
\begin{equation}
t=\sqrt[3]{\frac{12 \pi \rho_0 r_0^5}{\cal L} Y},
\end{equation}
where $Y$ is given by
\begin{equation}\label{kingsol}
Y=\left\{\begin{array}{l} 
\frac{1}{8} x (A^3+A/2)-\frac{1}{4} {\rm arcsinh}(x) B \\[1mm]
\frac{1}{2} \left( \frac{2}{3} x C -{\rm arctan(x)} A^2 \right) \\[1mm]
\frac{1}{2} {\rm arcsinh}(x) C - \frac{3}{4} x A \end{array}\right\}
{\rm for} \;3 \beta = \left\{ \begin{array}{l} 1 \\[1mm] 2 \\[1mm] 3 \end{array}\right. ,
\end{equation}
with
\begin{eqnarray*}
A&=&\sqrt{1+x^2}, \\
B&=&3/4\speco{19}{+}x^2, \\
C&=&3/2+x^2, \,\mathrm{and}\\
x&=&r/r_0.
\end{eqnarray*}
\begin{figure}[tb]
\centering
 \rotatebox{-90}{\resizebox{!}{.9\hsize}{\includegraphics{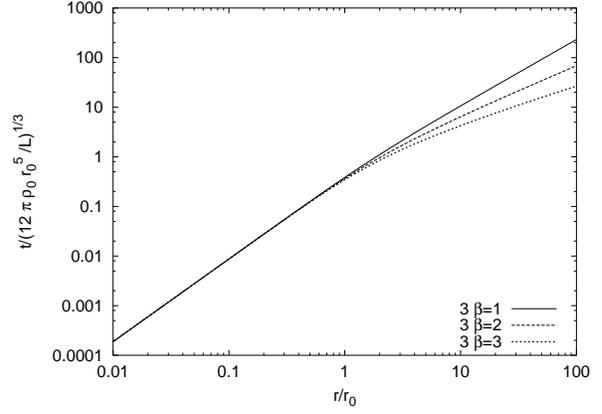}}}
 \caption{\label{king}\small Propagation of a blastwave powered by a jet into a King type
	atmosphere with $\beta$ as indicated above.}
\end{figure}
The solutions from Eq.~(\ref{kingsol}) are plotted in Fig.~\ref{king}. It is evident
that the effect of a King atmosphere, which is a smooth transformation from 
a constant density profile to a decline with exponent $-3\beta$, transforms the corresponding
power law solutions smoothly into one another. For $\beta=2/3$, which corresponds to
a $\kappa=\speco{20}{-}2$ power law solution for large $x$, the shell reaches constant
velocity at infinity. For steeper density declines, the shell accelerates.
In that case it is Rayleigh-Taylor unstable, and considerable mixing with the
cocoon can be expected, as already pointed out by \citet{HRB98}.
In the same manner as for the power law case, one can compute $r_1$ for the King 
density distributions. For example, in the $\beta=1$ case $r_1$ is given by the following
formula:
\begin{equation}
\eta \epsilon^3 = \left(\frac{3}{4}\right)^4 \left(\frac{r_\mathrm{j}}{r_0}\right)^4
	\left(1+x_1^2\right)^{-9/2} D(x_1)^{-6},
\end{equation}
where
\begin{eqnarray*}
x_1&=&r_1/r_0,\\
D(x_1)&=&E(x_1)^{-2/3} F(x_1),\\
E(x_1)&=&{\rm arcsinh}(x_1) C(x_1)-\frac{3}{2} x_1 A(x_1),\,\mathrm{and}\\
F(x_1)&=&x_1 {\rm arcsinh}(x_1) - x_1^2/A(x_1).
\end{eqnarray*}
$r_1$ is plotted for three different core radii in Fig.~\ref{r1king}\amd{.}
This density profile tends towards $r^{-3}$ for large r. As predicted above, $r_1$ is finite
also in that density regime. 
\speco{22}{For $\eta>10^{-4}$ and $r_0>10r_\mathrm{j}$,
the result is not much different from the constant
density case.
But in the more extreme cases $r_1$ can be reduced considerably, 
e.g. for $\eta=10^{-6}, r_0=2r_\mathrm{j}$, $r_1$ is reduced from 16 to 6 jet radii. 

Concerning the propagation law, the constant density formula approximates the King solution to 
10~\% up to approximately
one core radius for the shallowest density profile and half a core radius for the
steepest one (Fig.~\ref{king}).

Summarising, the constant density formulas can be used to approximate the true solutions
if $r_1$ is not much bigger than the core radius, if the density contrast is above $10^{-4}$,
and the density profile steepens not much more than into a $r^{-2}$ power law.}


\section{Discussion}
\label{disc}
A very interesting feature of the above results is the blastwave phase.
Since the blastwave is faster at early times, it paves the way for the jet,
effectively increasing the jet propagation velocity.
Later, the jet head is faster than the blastwave equation predicts. 
Therefore, an upper limit for jet ages is given by 
Eq.~(\ref{blastexp}).
\speco{1}{This can be used to establish a lower limit for $\eta$.
For very light jets, one can regard $r_1$ as a typical extention.
The time for a jet to reach the distance $r_1$ in a constant density atmosphere is:
\begin{equation}\label{time_estimate}
t(r_1)=\frac{1}{2 \sqrt[3]{5}} \eta^{-3/4} r_\mathrm{j} / v_\mathrm{j}.
\end{equation}
With the observational constraints cited in Sect.~\ref{obconst},
one obtains a new lower limit for the density contrast in extragalactic jets: $\eta > 10^{-7}$.} 


It was shown that the aspect ratio of the bow shock depends on the source size.
It is close to unity in the blastwave dominated phase, and starts to increase
soon after a critical size ($r_1$) is reached. This increase will come to an end
when the sideways expansion velocity drops to the sound speed.
\genco{1}{\citet{Saxea02} have computed similar models with density contrasts down to 
$\eta=10^{-4}$. The simulation results are generally
comparable in the regions where the simulation parameters 
overlap. The aspect ratio at the end of their $\eta=10^{-4}$ 
simulation is also close to 1.2.}
The aspect ratio can be used to constrain density ratios.
For example, the aspect ratio of 1.2 for the bow shock in Cygnus~A \citep{Sea01}
together with its extension of \amd{roughly} $120\, r_\mathrm{j}$ \citep{CarBar96}
are clearly inconsistent with $\eta \ge 10^{-2}$. Probably, $\eta$ should be below
$10^{-3}$. With cluster temperatures of \amd{approximately} $10^8$ K \citep{Sea01},
resulting in probable external Mach numbers of a few hundred, one derives
a limiting aspect ratio of about 2, which the source has not yet reached.
However, the jet clearly acts on the bow shock, and therefore one can check 
Eq.~(\ref{outbreak}), which is fulfilled if $\eta>3\times10^{-6}$.
Another result was that low cocoon width to bow shock width ratios can only
be achieved in low internal Mach number jets.
\speco{25}{This result is depends crucially on the closed left boundary,
which is essential in order to keep the internal energy inside of the 
computational domain.
Other studies have shown that the cocoon can be small even in high Mach number
jets for an open boundary on the injection side \citep{CarvOdea02a,Saxea02}.} 
Applying Eq.~(\ref{cocwid}) results in a probable range of 
$\eta \in [10^{-5};10^{-3}]$ for \omi $M \in [2;6]$ \cl{for the jet in Cygnus~A}.

With the general solution of the spherical force balance, one can also
predict how the results achieved here change in a nonhomogeneous 
density distribution. It was shown that -- for a King profile --
the blastwave phase \omi \amd{is} a little shorter for very small core radii.
But in most real cases, the blastwave phase will remain up to similar 
source sizes as in the constant density case.
This will be similar in all profiles with a constant density in the central region.
\genco{5}{Preliminary results from simulations with a King density distribution
support this conclusion.}

\genco{4}{The presented results are subject to two potentially important limitations:
The axisymmetry and the left hand boundary.
3D simulations of light jets have shown that big vortices are unlikely
to survive \citep{TJR01}. One should therefore expect turbulence on smaller scales in the 3D case.
The jet beams should be less disturbed, and might be able to sustain
higher Mach numbers. So, they might influence the bow shock earlier than
the 2D jets. However, since the 2D jets show this interaction already shortly
after reaching the theoretical $r_1$, the difference should not be severe in that respect.
The left boundary is an obstacle for material that would otherwise leave the grid there,
and interact with the counterjet. This should change the appearance of the cocoon 
in regions around $Z=0$. Both these issues will be addressed in a future paper
where a 3D simulation with bipolar (back-to-back) jets will be presented.}

By now, more and more X-ray bubbles are detected around radio sources,
mainly in CD galaxies 
\citep[e.g. ][ and references therein]{Sokea02}.
Such bubbles can be interpreted as \amd{gas contained within} the bow shock 
\amd{(or the bow wave, if the shock has relaxed)} from the jet. 
The theory presented in this work should be applicable 
to some of those sources. 
\speco{26}{The typical bubble size for which one should 
expect supersonic bow shocks (homogeneous atmosphere) is given by:
\begin{equation}\label{susobo}
r=24 \left[\frac{0.1 \;\mathrm{cm}^{-3}}{n_0}
\frac{\cal L}{10^{46}\mathrm{erg/s}}
\left(\frac{1000 \;\mathrm{km/s}}{c_\mathrm{s}}\right)^{3}\right]^{1/2}
\hspace{-2mm}\mathrm{kpc},
\end{equation} 
for typical values of the sound speed $c_\mathrm{s}$ and number density $n_0$
in the external medium. We should therefore expect no strong bow shocks around jet sources
with tens of kpc or more in diameter \citep[compare e.g. in Hydra~A,][]{Nulsea02}. 
This conclusion is not changed if one considers 
elongated bubbles, since the increased volume due to the elongation can only decrease 
the pressure inside the bubble. Hence, Eq.~\ref{susobo} gives an upper limit.
This means that the main energy transfer from radio sources 
to the gas of a galaxy cluster happens at early times in the jet evolution.
The power input into the cluster gas is 30~\% of the total jet power
for the constant density case.
This follows from integration of the energy gain at the bow shock. 
A strong jet source therefore has delivered 
\begin{eqnarray}
E_\mathrm{gas}&=&10^{60} \mathrm{erg} \left(\frac{\cal L}{10^{46}\,\mathrm{erg/s}}\right)^{3/2}
	\left(\frac{v_\mathrm{j}}{c/3}\right)^{-5/2}\nonumber \\
	& &\times \left(\frac{\eta}{10^{-4}}\right)^{-5/4}
	\left(\frac{n_0}{0.1\,\mathrm{cm}^{-3}}\right)^{-1/2}
\end{eqnarray}
to its surroundings
when its bow shock reaches $r_1$. This is already a considerable fraction of the
total energy of the gas in a typical galaxy cluster. Cooling flows can therefore 
easily be stopped by a powerful jet. One such episode every 1000~Myr 
could balance the energy loss and keep the cluster temperature in the observed keV
range.} 	 

\genco{2}{The results suggest that spherical bubbles should be found around most of the 
extragalactic jet sources at early evolutionary stages. The classes of 
compact symmetric objects (CSO) and GHz peaked sources (GPS) are  
associated with such sources.
Unfortunately, the bubbles are quite small. In order to resolve a region of a few kpc 
in diameter with Chandra, the source has to have a redshift considerably below one,
where few objects where found \citep{SneSchi02}. The hot ($10^9$~K) 
bubble would radiate in the X-rays
at about $10^{41}$~erg/s, hard to discern from a quasar background.

What about the radio morphology?
As was shown above, young jets in the bubble phase are often disturbed or disrupted.
This is also true for CSO/GPS sources \citep{Sakea02}. They are e.g. much more asymmetric 
than large scale jets. An old puzzle with CSO/GPS sources was if they were stuck within their
galaxy because of an especially dense interstellar medium (ISM) or because they are young.
With the results obtained above, one can exclude the first possibility.
Let us consider for example a jet with a power of $10^{47}$~erg/s propagating at near 
light speed through an incredibly dense ISM, $10^8$ times denser than the jet.
For the first kpc, a typical core radius, it would take the jet head about 
30~Myr, if propelled by the direct thrust alone. Such a jet was probably in the 
bubble phase for most, or maybe all, of its lifetime. 
The upper limit on its age from Eq. (\ref{blastexp}) turns out to be 
0.3~Myr, even for an ISM density of 100~$\mathrm{cm}^{-3}$. 
Therefore, the probability of finding a source
at a small angular size is not increased by considering dense environments.
Recent age determinations by spectral aging methods and VLBI kinematics of hotspots
strongly argue for young source ages of CSOs \citep{Con02}.}

The KH instability at the contact discontinuity is ubiquitous. This is especially
interesting since the same behaviour was found for $\eta=10^{-1}$ at highest 
resolution \citep{mypap01a}. The KH instability entrains shocked external medium into the jet cocoon,
growing and accelerating it towards the center. This mechanism seems to be a promising
candidate for bright X-ray features or the production of emission line regions, where 
the densities are higher \citep{Bickea00}: the gas is accelerated
and pulled down inside the cocoon, where it can radiate by recombination or reradiation
of quasar light. Given the high energy content of the radio lobes, one could also 
imagine that they \amd{somehow} transfer \omi energy to the entrained  gas.
Since the cocoon plasma is quite exotic, the classical heat conduction formulas
are probably not appropriate. Therefore, exact computations are beyond the scope of this work.

Concerning the beams, one can fairly say that those non-relativistic jets without significant
magnetic field cannot travel to large distances without being disrupted.
One needs a magnetic field in order to keep some of
the cocoon around the beam, and to damp the KH instabilities in the beam.
Alternatively, they would always look very disturbed, morphologically, until the bow shock
had swept enough material away, so they could effectively travel into a less dense medium.
Since bow shock detections point clearly in the direction of \speco{27}{strong} density contrast
\cl{(i.e. $\eta \ll 1$)},
a solution to this problem can be expected from further research.

\begin{acknowledgements}
This work was supported by the 
Deutsche Forschungsgemeinschaft
(Sonderforschungsbereich 439).
\amd{Thanks to the advice of the referee, Ian Tregillis,} 
\amd{this paper has been greatly improved.}
\end{acknowledgements}

\bibliographystyle{apj}
\bibliography{Krause.bib}

\end{document}